\documentstyle[12pt,aasms4,psfig]{article}

\def\etal{{\sl et al.}}

\slugcomment{Submitted to The Astrophysical Journal (Letters)}

\lefthead{Castro-Tirado \etal~ }
\righthead{Optical/IR counterpart to GRB 980703}

\begin{document}

\title{The optical/IR counterpart of the 3 July 1998 gamma-ray burst and 
its evolution\footnotemark}
\footnotetext{
Based on observations collected at the 
Observatorio del Teide, operated by the Instituto de
Astrof\'{\i}sica de Canarias (IAC); at the German-Spanish 
Astronomical Center, Calar-Alto, operated by the 
Max-Planck-Institut f\"ur Astronomie, Heidelberg, 
jointly with the Spanish National Commission for 
Astronomy; at the U. S. Naval Observatory, 
and with the Nordic Optical Telescope,
operated on the island of La Palma jointly by Denmark, 
Finland, Iceland, Norway and Sweden, in the Spanish
Observatorio del Roque de los Muchachos of the IAC.
}

\author{A. J. Castro-Tirado$^{2,3}$, M. R. Zapatero-Osorio$^4$,}

\author{J. Gorosabel$^2$, J. Greiner$^5$, J. Heidt$^6$, D. Herranz$^7$,}

\author{S. N. Kemp$^4$, E. Mart\'{\i}nez-Gonz\'alez$^7$, A. Oscoz$^4$,}

\author{V. Ortega$^4$, H.-J. R\"oser$^8$, C. Wolf$^8$,}

\author{H. Pedersen$^9$, A. O. Jaunsen$^{10}$, H. Korhonen$^{11}$,}

\author{I. Ilyin$^{12}$, R. Duemmler$^{12}$, M. I. Andersen$^{11}$,}

\author{J. Hjorth$^{9,13}$, A. A. Henden$^{14}$, F. J. Vrba$^{15}$,} 

\author{J. W. Fried$^{8}$, F. Frontera$^{16,17}$, and L. Nicastro$^{18}$}

\bigskip 
\bigskip 

\affil{$^2$ Laboratorio de Astrof\'{\i}sica Espacial y F\'{\i}sica 
              Fundamental (LAEFF-INTA), 
              P.O. Box 50727, E-28080 Madrid, Spain}

\affil{$^3$ Instituto de Astrof\'{\i}sica de Andaluc\'{\i}a (IAA-CSIC), 
              P.O. Box 03004, E-18080 Granada, Spain}

\affil{$^4$ Instituto de Astrof\'{\i}sica de Canarias, 
              E-38200, La Laguna, Tenerife, Spain}

\affil{$^5$ Astrophysikalisches Institut Potsdam, D-14482 Potsdam, Germany} 

\affil{$^6$ Landessternwarte Heidelberg, K\"onigstuhl, D-69117 Heidelberg, 
       Germany}

\affil{$^7$ Instituto de F\'{\i}sica de Cantabria (IFCA, CSIC-UC), 
       Santander, Spain}

\affil{$^8$ Max-Planck-Institut f\"ur Astronomie, Koenigstuhl, D-69117 
       Heidelberg, Germany  }

\affil{$^9$ Copenhagen Astronomical Observatory, Juliane Maries Vej 30, 
              DK-2100 Copenhagen, Denmark}   

\affil{$^{10}$ Institute for Theoretical Astrophysics, Oslo}

\affil{$^{11}$ Nordic Optical Telescope, La Palma, Spain}

\affil{$^{12}$ Astronomy Division, University of Oulu, P.O. Box 333, FIN-90571
              Oulu, Finland}

\affil{$^{13}$ Nordic Institute for Theoretical Physics (NORDITA), 
        Blegdamsvej 17, DK-2100 Copenhagen, Denmark}

\affil{$^{14}$ U.S.R.A./U. S. Naval Observatory, Flagstaff Station, Flagstaff,
       Arizona 86002, USA}

\affil{$^{15}$ U. S. Naval Observatory, Flagstaff Station, Flagstaff,
       Arizona 86002, USA}

\affil{$^{16}$ Istituto Tecnologie e Studio Radiazioni Extraterrestri, CNR, 
        Via P. Gobetti, I-40129 Bologna, Italy}

\affil{$^{17}$ Dipartimento di Fisica, Universit\`a di Ferrara, I-44100 
        Ferrara, Italy}

\affil{$^{18}$ Istituto di Fisica Cosmica ed Applicazioni 
        all$^{\prime}$Informatica, CNR, Via U. La Malfa 153, 
        I-90146 Palermo, Italy}

\begin{abstract}

  We imaged the X-ray error box of GRB 980703, beginning 22.5 hours
  after the $\gamma$--ray event, in both the optical R and near-infrared H
  bands. A fading optical/IR object was detected within the X-ray error
  box, coincident with the variable radio source reported by Frail
  et al. (1998a), who also detected the optical transient independently of us.
  Further imagery revealed the GRB host galaxy, with R = 22.49 $\pm$ 0.04
  and H = 20.5 $\pm$ 0.25, the brightest so far detected.  When excluding
  its contribution to the total flux, both the R and H-band light curves
  are well-fit by a power-law decay with index $\alpha$ $\simeq$
  1.4. Our data suggest an intrinsic column density in the host
  galaxy of
  $\sim$ 3.5 $\times$ 10$^{21}$ cm$^{-2}$ which indicates the existence of
  a dense and rich-gas medium in which the GRB occurred, thus supporting
  the hypernova model scenarios.  

\end{abstract}


\section{Introduction}

Gamma--ray bursts (GRBs hereafter) are brief flashes of cosmic high energy
photons. With the advent of the BeppoSAX satellite, it has been possible
for the first time to perform deep multiwavelength searches hours after the
event, allowing detection in a few cases of transient emission associated with
the GRB.  The modelling of the afterglow decay is well understood but their
central engines remain enigmatic.

GRB 980703 was detected on 1998 July 3.182465 UT by the
Burst And Transient Source Experiment (BATSE) on board the Compton
Gamma-Ray Observatory (Kippen et al. 1998), the All-Sky Monitor (ASM) on the 
Rossi X-ray Timing Explorer (RXTE) (Levine, Morgan and Nuno 1998) 
and the GRB Monitor on BeppoSAX (Amati et al. 1998).  The X--ray flux
as seen by RXTE (2--10 keV) began to rise 18 s before the BATSE flux trigger
(\#6891) and the GRB position was constrained to a 12$^{\prime}$ $\times$
12$^{\prime}$ diamond-shaped error box (Smith et al. 1998). The duration as
seen by BATSE was about 400 s with two well distinguished peaks, each of
them lasting about 100 s, reaching a peak flux (50--300 keV, integrated
over 1 s) of 2.58 $\pm$ 0.12 photons cm$^{-2}$ s$^{-1}$, and a
fluence ($\geq$ 20 keV) of 6.2 ($\pm$ 0.4) $\times$ 10$^{-5}$ erg
cm$^{-2}$ (Kippen et al. 1998).

After 22 hours, the BeppoSAX X-ray satellite was reoriented towards the
burst location, and one previously unknown X--ray source (labelled 1SAX
J2359.1+0835) was detected wi\-thin the error box provided by RXTE (Galama
et al. 1998a).
The observed flux (2-10 keV) during the first 8,200-s was found to be 7.3
$\times$ 10$^{-13}$ erg cm$^{-2}$ s$^{-1}$, decreasing in the following
30,000 s to 4.8 $\times$ 10$^{-13}$ erg cm$^{-2}$ s$^{-1}$, implying a
decay index $\alpha_{X}$ of 1.05$^{+0.23}_{-0.29}$ (Vreeswijk et al. 1998), 
asuming a power-law decay of the flux F $\propto$ t$^{-\alpha_{X}}$. This 
decay supports its identification as the X-ray afterglow of GRB 980703.

A comparison among the R-band frames acquired on July 4 and July 6 allowed
us to search for variable objects within a 2$^{\prime}$ radius circle
centered at the position of the variable X-ray source presumably related to
the GRB. The position of the optical transient
reported by Zapatero-Osorio et al. (1998a,b), is consistent with the 
position of the source discovered independently of us by Frail et al. (1998a).

Djorgovski et al. (1998a) discovered a host galaxy to GRB 980703 which 
lies at a redshift z = 0.966 (Djorgovski et al. 1998b). This implies that 
the isotropic energy release was $\sim$ 10$^{53}$ erg assuming a standard 
Friedman model cosmology with 
$H_0$ = 65 km s$^{-1}$ Mpc$^{-1}$ and $\Omega_0$ = 0.2.
Here we present the optical and near-infrared observations
carried out by us soon after the occurrence of the high-energy event.

\section{Observations and Data Reductions}

The observations in the Johnson R-band were obtained on July 4, 6 and 7
using the 0.8-m telescope located at the Observatorio del Teide. The CCD
used was a Thomson 1024 $\times$ 1024 mounted on the Cassegrain focus and
providing a field of view of 7.4$^{\prime}$ $\times$ 7.4$^{\prime}$.  On 
the first night (July 4), the telescope was pointed at four different
positions (an offset of 3 arcmin was applied both in RA and DEC) covering 
an area of 109.5 arcmin$^2$ around the RXTE central coordinates. In this  
way about 90\% \ of the diamond-shaped error box of RXTE (Smith et al. 1998) 
was surveyed in the R-filter.  The optical transient reported in this
paper appeared in three of the four frames. On the following two observing
nights, we pointed toward 1SAX J2359.1+0835, which was proposed to be the 
candidate for the X-ray afterglow of GRB 980703. Exposure times were
typically 1800 s per image. Weather conditions during the three nights were
photometric, however no standard star observations were
obtained. The seeing ranged between 1.5 and 1.9 arcsec.  Raw frames
were processed with standard techniques using routines within the 
IRAF\footnote{IRAF is
  distributed by National Optical Astronomy Observatories, which is
  operated by the Association of Universities for Research in Astronomy,
  Inc., under contract with the National Science Foundation.} 
(Image Reduction and Analysis Facility) environment, which included 
bias substraction, flat-fielding (using flat 
images obtained during twilight) and correction for bad pixels by 
interpolation with values from the nearest-neighbour pixels. A common area 
of 9.8 arcmin$^{2}$ was covered during all three observing nights. 
Instrumental magnitudes were obtained using the DAOPHOT package 
(Stetson 1992, and references therein), 
and were transformed into
Cousins R photometry adopting calibrated data for this field provided by
Rhoads et al. (1998), who used observations of several Landolt
(1992) photometric standard stars. Uncertainties quoted for our R measurements
in Table 2 come from quadratic combination of the dispersion in the
photometric calibration (typically $\pm$0.03 mag) and the intrinsic 
errors of the instrumental magnitudes. The limiting R magnitude is 22.7 in
all images.  Additional R-band data were obtained with the Nordic Optical 
Telescope (NOT) on July 7.163 UT (and the Standby Camera, StanCam) and 
July 20.172 UT (and the Andaluc\'{\i}a Faint Object Spectrograph Camera, 
ALFOSC).

\placetable{tab1}

\placefigure{fig1}

\placetable{tab2}

\placefigure{fig2} 
 
Near-simultaneous observations were obtained in the H-band with the 3.5-m
telescope (equipped with Omega) at Calar Alto (CAHA).  
Omega is a near-IR detector with
a 1024 $\times$ 1024 pixel HgCdTe array. The image scale is
0.4$^{\prime\prime}$/pixel giving a field of view of 6.8$^{\prime}$
$\times$ 6.8$^{\prime}$ (Bizenberger et al. 1998).  We obtained H-band
images beginning 22 h 38 min after the event (July 4.125 UT).
The field was centered on the initial GRB location provided by RXTE. A
10$^{\prime\prime}$ dithering around the GRB coordinates was performed, thus
obtaining five frames (120-s each) in order to determine and subtract the
background from each of the individual images. 

Another set of images were obtained at CAHA two days after, following 
a different pattern (9 frames, 120-s
each). A further observation was obtained on Aug 9 at the 3.5-m CAHA. A
3$\times$3 mosaic was performed, with each position being observed
28 s. The mosaic was repeated 14 times, thus giving a total
on-source exposure of 3538-s.  The H-band calibration is based on the
observations of the nearby UKIRT (UK Infrared Telescope) faint standard 
FS 3 (Cassali \& Hawarden 1992; Hunt et al. 1998) under photometric 
conditions on Aug 9, just before observing the GRB 980703 field. 

Additional observations were also obtained with the U. S. Naval 
Observatory (USNO) Flagstaff Station 1.55-m telescope and the Infrared 
Camera (IRCAM). IRCAM is a near-IR camera with
a 256 $\times$ 256 pixel Rockwell/NICMOS HgCdTe array. The image scale is
0.54$^{\prime\prime}$/pixel giving a field of view of 2.3$^{\prime}$
$\times$ 2.3$^{\prime}$. Imaging in the H-band was performed on July 
4.458 and 8.459 under photometric conditions (see Henden et al. 1998a,b 
for the preliminary results).
 
All the H-band data points were obtained with the same secondary standard, 
a star at 12$^{\prime\prime}$ W and 10$^{\prime\prime}$ S from the OT, with
H = 15.52 $\pm$ 0.05. 
Photometry was performed by means of SExtractor 2.0 (Bertin and Arnouts
1996), making use of the corrected isophotal magnitude, which is very
appropriate for star-like objects.  A log of these optical/IR observations 
is given in Table 1. Magnitudes are given in Table 2.

\section{Results and Discussion}

As we have already mentioned, a comparison among the R-band frames 
acquired on July 4 and July 6 allowed
us to search for variable objects within a 2$^{\prime}$ radius circle
centered at the position of the variable X-ray source presumably related to
the GRB.  With the exception of one source (Fig. 1), no object brighter
than R $\sim$ 22.5 was seen to vary by more than 0.2 mag.  Astrometry for
this candidate was achieved by the triangles fitting method using the APM
(Automatic Plate Measuring machine) Sky Catalogue. Several stars 
close to it were identified and they served as
a reference for the astrometric calibration. The optical transient was
found at R.A.(2000) = 23$^{h}$59$^{m}$06.7$^{s}$, Dec(2000) = +08$^{\circ}$
35$^{\prime}$ 07$^{\prime\prime}$ ($\pm$ 1$^{\prime\prime}$)
(Zapatero-Osorio et al.  1998a,b), consistent with the position of the
source discovered independently of us by Frail et al. (1998a).

We find that, about 22.5 h after the burst, the optical transient had R
= 21.00 $\pm$ 0.09, H = 17.30 $\pm$ 0.10.  The light--curves in both the R
and H-bands are shown in Fig. 2. We have included the H-band data from Bloom 
et al. (1998b). In the R-band, the source declined in
brightness following a power law, with the flux F $\propto$ t$^{-\alpha}$
with $\alpha$ = 1.39 $\pm$ 0.3 ($\chi^{2}$/dof = 0.09), a value that has
been obtained when excluding the contribution of a constant source (R =
22.49 $\pm$ 0.04), the brightest host galaxy detected so far. We note that 
the low value of the chi-squared-reduced value is due to the fact that the 
errors in the data points have been overestimated. 
Regarding the near-infrared data, a pure power-law fit with index
$\alpha = 1.37 \pm 0.24$ ($\chi^{2}$/dof = 0.1) is consistent with the
first four H-band measurements, once we have excluded the host galaxy 
contribution (H = 20.5 $\pm$ 0.25). These two values are consistent with
$\alpha = 1.17 \pm 0.25$ reported by Bloom et al. (1998b).  

The emission from X-rays to radio energies that follows the gamma-ray event 
is known
as the afterglow. This is caused by the relativistic blast wave in which
the emitted synchrotron radiation arises from relativistic electrons with
high Lorentz factors $\gamma_{e}$. The electrons follow a power-law
distribution N($\gamma_{e}$) $\propto$ $\gamma_{e}^{-p}$ above some minimum
Lorentz factor $\gamma_{m}$, i.e. for frequencies $\nu$ $>$ $\nu_{m}$, with
$\nu_{m}$ being the synchrotron peak frequency, following Sari et al.
(1998).

On July 4.1 UT, the equivalent spectral slope $\beta$ (with F$_{\nu}$
$\propto$ $\nu^{\beta}$) between the R-band and the X-rays is $\beta_{RX}$
= $-$0.7 $\pm$ 0.1, similar to the spectral indexes of other X-ray
afterglows in which $\beta_{RX}$ $\simeq$ $-$0.5 to $-$1.0 
(Galama et al. 1998b, Halpern et al. 1998). 
Our near-simultaneous R- and H-band measurements on July 4.12 UT, R-H
= 3.70 $\pm$ 0.14, implies an optical-to-near-infrared index of
$\beta_{HR}$ $\simeq$ $-$2.8 (or $\simeq$ $-$3.1 excluding the contribution
of the host galaxy), indicative of a reddened spectrum. Thus, the
broad-band spectrum from the near-infrared to X-rays deviates from a
single power-law that should have been seen once the synchrotron break has
passed the optical band. Blast wave models (Sari et al. 1998 and
references therein) predict much flatter optical spectra in these bands
with an optical-to-near-infrared index $\beta_{HR}$ in the range $-$0.5
to $-$1.5. According to the dust maps of Schlegel et al.  (1992), the dust
column density in the direction of GRB 980703 implies E(B-V) = 0.06, i.e.
A$_{V}$ = 0.2. This cannot explain the observed steep spectrum.  We then
consider the presence of extra absorption. In fact, one possibility to
straighten the observed slope is by assuming intrinsic absorption.  Let us
determine the amount of intrinsic absorption needed in order to de-redden
the spectrum to get $\beta_{HR}$ $\simeq$ $-$0.6 (and $\beta_{RX}$ $\simeq$
$-$1 as expected in the cooling regime), close to the observed value of 
$\beta_{HR}$ $\simeq$ $-$0.5 in GRB 970508 (derived from Fig.2 of Palazzi et 
al.  1998), a burst which is
considered to have had negligible absorption.  Assuming a galactic
extinction-law type (Savage and Mathis 1979) at the proposed host galaxy
which lies at a redshift z = 0.966 (Djorgovski et al.  1998b), we find that
E(B-V) $\simeq$ 0.7 (i.e.  A$_{V}$ = 2.2) is required, a value different than
A$_{V}$ = 0.9 found by Bloom et al. (1998b). This corresponds to
an intrinsic column density of $\sim$ 3.5 $\times$ 10$^{21}$ cm$^{-2}$.

If we assume that by the time of our first observations, the peak frequency
$\nu_{m}$, has already passed the H-band, and assuming
adiabatic evolution of the GRB remnant, the decay of the flux after the
passage of $\nu_{m}$ would go as F$_{\nu}$ $\propto$ t$^{3(1-p)/4}$ =
t$^{-\alpha}$ with $\alpha$ $\simeq$ 1.3 as observed in the R-band (better
sampled than H), giving $p$ = 2.7, similar to GRB 971214.

The derived intrinsic column density indicates the presence of dense medium 
rich in
gas, surrounding the GRB. This is supported by the proposed identification
of the host galaxy as a compact, star forming galaxy (Djorgovski et al.
1998b) and once more, gives strength to the ``failed'' supernova 
(Woosley 1993) 
or microquasar (Paczy\'nski 1998) model scenarios. Both types of objects 
are thought to be found in dense, star-forming regions.

\acknowledgments

We are very grateful to J. Bloom for having provided one of the H-band 
images obtained at the W. M. Keck telescope.  Also to J. L. Sanz, N.
Ben\'{\i}tez, T. Broadhurst and J. Silk, for lending part of their
observing time.  This work has been (partially) supported by Spanish CICYT
grant ESP95-0389-C02-02. JG is supported by the Deut\-sches Zentrum f\"ur
Luft- und Raumfahrt (DLR) GmbH under contract No. FKZ 50 QQ 9602\,3.

\clearpage

\begin{table}[bt]
\begin{tabular}{l|c|c|c|c}
\multicolumn{5}{c}{Table 1. Journal of the GRB 980703 observations }\\
\hline
\hline
        &           &           &        &              \\
Date    & Time (UT) & Telescope & Filter & Integration  \\
        &           &           &        & time (s)     \\
\hline
04 July 98  & 02 h 57 min  & 0.8 IAC    & R  & 3 x 1800 \\
06 July 98  & 04 h 03 min  & 0.8 IAC    & R  & 4 x 1800 \\
07 July 98  & 03 h 54 min  & 2.56 NOT    & R  &   600    \\
07 July 98  & 04 h 55 min  & 0.8 IAC    & R  & 2 x 1800 \\
20 July 98  & 04 h 07 min  & 2.56 NOT    & R  & 5 x  600 \\
\hline
04 July 98  & 03 h 00 min  & 3.5 CAHA   & H  &   600    \\
04 July 98  & 10 h 59 min  & 1.55 USNO  & H  &   240    \\
06 July 98  & 02 h 58 min  & 3.5 CAHA   & H  &  1080    \\
08 July 98  & 11 h 00 min  & 1.55 USNO  & H  &  2400    \\
09 Aug  98  & 03 h 15 min  & 3.5 CAHA   & H  &  3528    \\
\hline
\end{tabular} 
\end{table}

\clearpage

\begin{table}[t]
\begin{tabular}{l|c|c|c}
\multicolumn{4}{c}{Table 2. Optical and near-infrared magnitudes }\\
\hline
\hline
                  &           &           &            \\
Date of 1998      & Telescope & Magnitude & Ref.  \\
   (UT)           &           &           &             \\
\hline
July 4.123 & 0.8-m IAC   & R = 21.00 $\pm$ 0.09 & 1           \\
July 4.166 & 0.8-m IAC   & R = 21.04 $\pm$ 0.06 & 1           \\
July 4.188 & 0.8-m IAC   & R = 21.13 $\pm$ 0.05 & 1           \\
July 4.418 & 0.9-m Kitt Peak & R = 21.17 $\pm$ 0.12 & 2   \\
July 4.433 & 0.9-m Kitt Peak & R = 21.35 $\pm$ 0.11 & 2   \\
July 4.448 & 0.9-m Kitt Peak & R = 21.22 $\pm$ 0.10 & 2   \\
July 4.477 & 1.52-m Palomar  & R = 21.34 $\pm$ 0.2  & 2, 3  \\
July 5.482 & 1.52-m Palomar  & R = 21.8 $\pm$ 0.3  & 3 \\
July 6.137 & 0.8-m IAC   & R = 22.01 $\pm$ 0.11 & 1           \\
July 6.607  &10.0-m Keck  & R = 22.04 $\pm$ 0.2 & 2, 3  \\
July 7.163 & 2.56-m NOT   & R = 22.21  $\pm$ 0.09 & 1 \\
July 7.205 & 0.8-m IAC   & R = 22.6 $\pm$ 0.2   & 1         \\
July 20.172 & 2.56-m NOT & R = 22.49 $\pm$ 0.04  & 1           \\
July 25.01 & 6.0-m BTA   & R = 22.43 $\pm$ 0.08 & 4   \\
\hline
July 4.125 & 3.5-m CAHA  & H = 17.3 $\pm$ 0.1 & 1           \\
July 4.458 & 1.55-m USNO & H = 17.9 $\pm$ 0.3  & 1  \\
July 6.125 & 3.5-m CAHA  & H = 18.85 $\pm$ 0.2 & 1           \\
July 8.459 & 1.55-m USNO & H $>$ 19.4        & 5             \\
July 8.5764 &10.0-m Keck & H = 19.56 $\pm$ 0.13 & 6           \\
Aug  7.52   &10.0-m Keck & H = 20.3 $\pm$ 0.2  & 6           \\
Aug  9.135 & 3.5-m CAHA  & H = 20.5 $\pm$ 0.25  & 1            \\
\hline
\multicolumn{4}{l}{References: [1] This paper; [2] Rhoads et al. 1998;}\\
\multicolumn{4}{l}{[3] Bloom et al. 1998a; [4] Sokolov et al. 1998;}\\
\multicolumn{4}{l}{[5] Henden et al. 1998b; [6] Bloom et al. 1998b}\\
\end{tabular} 
\end{table}

\clearpage

\begin{figure}
\figurenum{1}


\caption{
A $2^{\prime} \times 2^{\prime}$ field containing  
      the 50$^{\prime\prime}$ error box radius for 1SAX J2359.1+0835 
      (Galama et al. 1998a), as imaged at the IAC-80 on 4, 6 and 1998 July 7
      in the R-band. An optical transient ({\it arrow}) was reported 
      (Zapatero-Osorio et al. 1998a,b) at the position of the variable radio
      (and optical) source proposed by Frail et al. (1998). 
      Limiting magnitude is R $\sim 22.7$ in all the images. 
\label{fig1}
}
\end{figure}


\begin{figure}
\figurenum{2}


\caption{
The light curve of GRB 980703 in the R and H-bands. 
Based on our observations ({\it filled circles}) and other data 
({\it diamonds}) obtained from the literature (see Table 2).
The dot-dashed line is the contribution of the host 
galaxy, with R = 22.49 $\pm$ 0.04 and H = 20.5 $\pm$ 0.25. 
The long-dashed line is
the contribution of the OT, following F $\propto$ $t^{-\alpha}$ 
with $\alpha$ = 1.39 in R and $\alpha$ = 1.37 in H. The solid 
line is the total observed flux (OT + host galaxy), following 
F = F$_{0}$ $t^{-\alpha}$ + F$_{host}$. 
\label{fig2}
}
\end{figure}

\clearpage

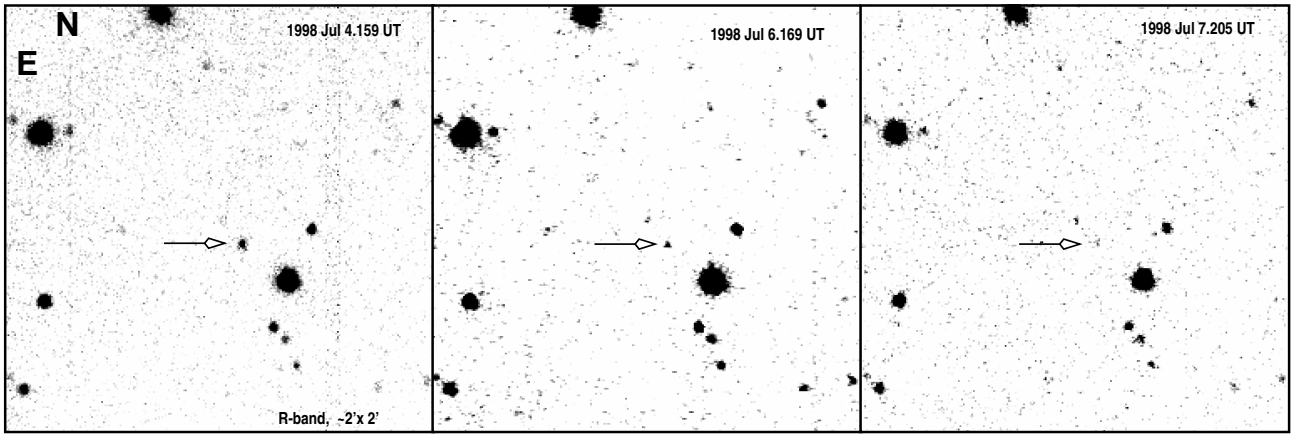
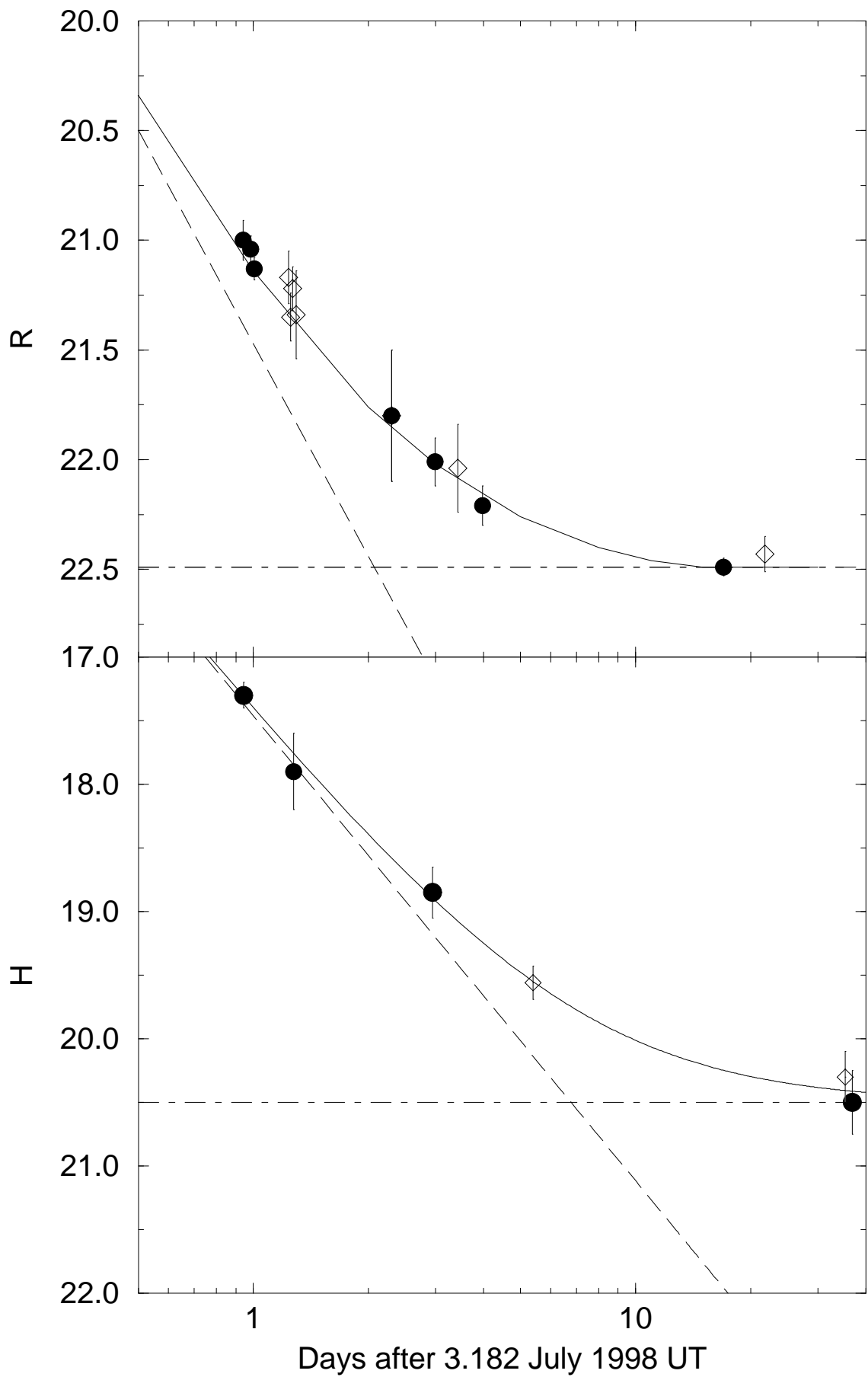
\end{document}